\documentclass[10pt]{iopart}

\usepackage{graphicx}  
\usepackage{epsfig}
\usepackage[numbers]{natbib}
\usepackage{subfigure}
\usepackage{color}
\usepackage{amsfonts}
\usepackage{hyperref}
\usepackage{ulem}
\usepackage{longtable}
\usepackage{lscape}
\usepackage{url}
\usepackage{doi}

\newcommand{\xv}{\textbf{x}}
\newcommand{\xvp}{\textbf{x}^{\prime}}  
\newcommand{\half}{\frac{1}{2}}

\begin{document}

\review[Chimera states]{Chimera states: Coexistence of coherence and incoherence in networks of coupled oscillators} 

\author{Mark J. Panaggio$^{1,2}$, Daniel M. Abrams$^{1,3,4}$}
\address{$^1$ Department of Engineering Sciences and Applied Mathematics, Northwestern University, Evanston, Illinois 60208, USA}
\address{$^2$Mathematics Department, Rose-Hulman Institute of Technology, Terre Haute, IN 47803, USA}
\address{$^3$ Department of Physics and Astronomy, Northwestern University, Evanston, Illinois 60208, USA}
\address{$^4$ Northwestern Institute on Complex Systems, Evanston, Illinois 60208, USA}
\ead{markpanaggio2014@u.northwestern.edu}
\begin{abstract}
A chimera state is a spatio-temporal pattern in a network of identical coupled oscillators in which synchronous and asynchronous oscillation coexist. This state of broken symmetry, which usually coexists with a stable spatially symmetric state, has intrigued the nonlinear dynamics community since its discovery in the early 2000s.  Recent experiments have led to increasing interest in the origin and dynamics of these states.  Here we review the history of research on chimera states and highlight major advances in understanding their behaviour.
\end{abstract}
% 02.30.Ks - Mathematical methods in physics - Delay and functional equations
% 05.45.Ra - Statistical physics, thermodynamics, and nonlinear dynamical systems - Coupled map lattices
% 05.45.Xt - Statistical physics, thermodynamics, and nonlinear dynamical systems - Synchronization; coupled oscillators
% 05.65.+b - Statistical physics, thermodynamics, and nonlinear dynamical systems - Self-organized systems
% 82.40.Bj - Physical chemistry and chemical physics - Oscillations, chaos, and bifurcations
% 82.40.Ck - Physical chemistry and chemical physics - Pattern formation in reactions with diffusion, flow and heat transfer 
% 87.85.dq - Biological and medical physics - Neural networks
% 89.75.-k - Other areas of applied and interdisciplinary physics - Complex systems 
% 89.75.Kd - Other areas of applied and interdisciplinary physics - Patterns

% 34C15 - Ordinary differential equations - Nonlinear oscillations, coupled oscillators
% 82C22 - Statistical mechanics, structure of matter - Interacting particle systems
% 34C23 - Ordinary differential equations - Bifurcation 
% 35B36 - Partial differential equations - Pattern formation
% 82C44 - Statistical mechanics, structure of matter - 	Dynamics of disordered systems 
%\submitto{\NL}
\maketitle

%************ Section ************%
\section{Background} \label{sec:background}
\begin{figure*}[bth!]
{
    \centering
	\includegraphics[width=1\textwidth]{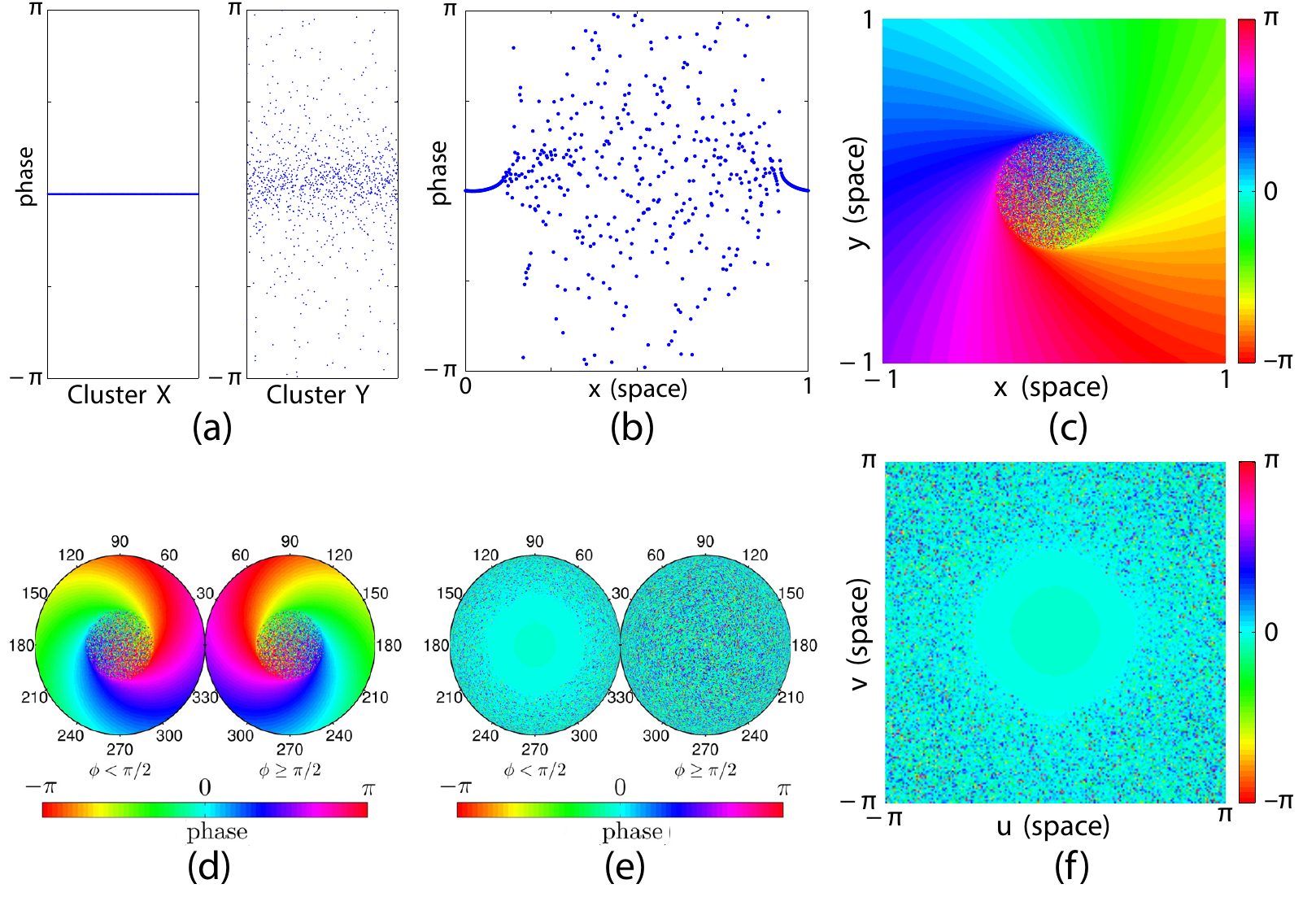} 
	\caption{Examples of chimera states. (a) Chimera state in a system of two point-like clusters.  (b) Chimera state in a one-dimensional periodic space (ring). (c) The incoherent region for a spiral chimera state on a two-dimensional infinite plane.  (d) `Spiral' chimera state on a two-dimensional periodic space (sphere).  (e) `Spot' chimera state on a two-dimensional  periodic space (sphere). (f) `Spot' chimera state on a two-dimensional periodic space (flat torus). } 
	\label{fig:examples}}
\end{figure*}

In Greek mythology, the chimera was a fierce fire-breathing hybrid of a lion, a goat and a snake.  In the nonlinear dynamics community, however, `chimera' has come to refer to a surprising mathematical hybrid, a state of mixed synchronous and asynchronous behaviour in a network of identical coupled oscillators (see \fref{fig:examples}).  

Until about ten years ago, it was believed that the dynamics of networks of identical {phase-oscillators ($d\theta_i/dt=\omega + \textrm{coupling}$)} were relatively uninteresting.  Whereas coupled \textit{non-identical} oscillators were known to exhibit complex phenomena including frequency locking, phase synchronisation, partial synchronisation, and incoherence, \textit{identical} oscillators were expected to either synchronise in phase or drift incoherently indefinitely. Then, in November 2002, Japanese physicist Yoshiki Kuramoto (already well-known for his paradigmatic model of synchronisation in phase oscillators \cite{Kuramoto1975,Pikovsky2003,Acebron2005,Strogatz2000}) and his collaborator Dorjsuren Battogtokh showed that the conventional wisdom was wrong \cite{Kuramoto2002}.  While investigating a ring of identical and non-locally coupled phase oscillators, they discovered something remarkable: for certain initial conditions, oscillators that were identically coupled to their neighbors and had identical natural frequencies could behave differently from one another.  That is, some of the oscillators could synchronise while others remained incoherent \cite{Kuramoto2002}.  This was not a transient state, but apparently a stable persistent phenomenon combining some aspects of the synchronous state with other aspects of the incoherent state\footnote{In many systems this state coexists with a stable fully-synchronised state---this long concealed its existence.}. Steve Strogatz later had the idea to dub these patterns ``chimera states'' for their similarity to the mythological Greek beast made up of incongruous parts \cite{Abrams2004}.

Early investigations of chimera states prompted many questions.  Were these patterns stable? Did they exist in higher dimensional systems? Were they robust to noise and to heterogeneities in the natural frequencies and coupling topology? Were they robust enough to be observable in experiments?  Could more complex patterns of asynchronous and synchronous oscillation also be observed? Could the dynamics of these patterns be reduced to lower dimensional manifolds? What are the necessary conditions for a chimera state to exist?

During the last decade, many of these questions have been answered.  We now know that, for certain systems, though they are stable as the number of oscillators $N \to \infty$, chimera states are actually very long lived transients for finite $N$. Although the basins of attraction for chimera states are typically smaller than that of the fully coherent state, chimera states are robust to many different types of perturbations. They can occur in a variety of different coupling topologies and are even observable in experiments. 

In this review, we will highlight some important results pertaining to chimera states since their discovery and explore potential applications of these unusual dynamical patterns.

%************ Section ************%
\section{What is a chimera state?}

Abrams and Strogatz defined a chimera state as a spatio-temporal pattern in which a system of \textit{identical} oscillators is split into coexisting regions of coherent (phase and frequency locked) and incoherent (drifting) oscillation.  On their own, neither of these behaviours were unexpected.  Both incoherence and coherence were well-documented in arrays of \textit{non-identical} coupled oscillators, but {complete incoherence and partial coherence were usually stable at different coupling strengths}. It was believed that coexistence was only possible due to heterogeneities in the natural frequencies. Nonetheless, Kuramoto and Battogtokh observed a chimera state when all of the oscillators were identical.  {They considered the system}:
\begin{equation}   \label{eq:kuramoto}
  \frac{\partial}{\partial t} \psi(x,t)=\omega(x) -\int G(x-x^{\prime})\sin(\psi(x,t)-\psi(x^{\prime},t)+\alpha)dx^{\prime}~. 
\end{equation} 
{with $\omega(x)=\omega$ for all $x$} \footnote{{There is some ambiguity in how the integral in equation \eref{eq:kuramoto} should be evaluated. One possibility is that the equation can be treated as an ``abbreviation'' for the discrete Kuramoto model (see equation \eref{discrete_kuramoto} with $\epsilon=1$). In this case, the integral is replaced by a sum over a countable number of oscillators. Alternatively, one can interpret equation \eref{eq:kuramoto} as if there were a distribution of oscillators at each point in space. In this case, the integral should be interpreted as $\int G(x-x')\int_0^{2\pi}\sin(\psi(x,t)-\psi'+\alpha)p(\psi',x',t)d\psi' dx'$ where $p(\psi',x',t)$ represents the probability distribution of the phases and satisfies the continuity equation (see equation\eref{eq:continuity}---in other words, \eref{eq:kuramoto} is no longer sufficient to describe the dynamics on its own).}}. Apparently, only non-local/non-global coupling (non-constant $G(x)$)  and non-zero phase lag $\alpha$ were required to observe coexistence of these divergent behaviors.  This result was particularly surprising because it occurred in regions of parameter space where the fully coherent state was also stable.  Thus, the symmetry breaking in the dynamics was not due to structural inhomogeneities in the coupling topology.  So where did this state come from?

%************ Section ************%
\section{A simple example}
\label{sec:simpleexample}
To see why chimera states are possible, it is instructive to consider the simplest system where they have been observed: a network with two clusters of $N$ identical oscillators \cite{Abrams2008}. Because they are identical and identically coupled, all oscillators are governed by the same equation,
\begin{eqnarray}\label{eq:gov}
  \frac{d \theta_i^{\sigma}}{dt} &=& \omega - \mu \left< 
      \sin ( \theta_i^{\sigma} - \theta_j^{\sigma} + \alpha ) 
    \right>_{j\in \sigma} - \nu \left<
      \sin ( \theta_i^{\sigma} - \theta_j^{\sigma^{\prime}} + \alpha ) 
    \right>_{j\in \sigma^{\prime}}~,
\end{eqnarray}
where $\mu$ and $\nu$ represent the intra- and inter-cluster coupling strengths respectively ($\mu > \nu > 0$), {$\sigma$ and $\sigma^{\prime}$ indicate clusters $X$ and $Y$ (or vice versa) and $\left<f(\theta_j^{\sigma})\right>_{j\in \sigma}$ indicates an average over cluster $\sigma$ (this is just $N^{-1} \sum_{j=1}^N f(\theta_j^{\sigma})$ for finite $N$). When $N \to \infty$, the phases in each cluster have a probability density function $p^{\sigma}$ and the cluster average $\left<f(\theta_j^{\sigma})\right>_{j\in \sigma}$ is defined as $\int_0^{2\pi} f(\theta') p^{\sigma}(\theta',t) d\theta'$.} These probability distributions must satisfy the continuity equation
\begin{equation} \label{eq:continuity}
\frac{\partial p^{\sigma} }{\partial t} + \frac{\partial }{\partial
\theta } ( p^{\sigma} v^{\sigma} ) = 0~,
\end{equation}
where $v^{\sigma}$ is the phase velocity given by equation \eref{eq:gov}, but with $\theta_i^{\sigma}$ replaced by a continuous $\theta$ and sums replaced by integrals:
\begin{equation} \label{eq:v_Def}
  \hspace{-1.8cm}
  v^{\sigma} (\theta, t) = \omega - \mu \int_0^{2\pi}
    \sin ( \theta - \theta' + \alpha ) p^\sigma(\theta',t) 
  d \theta' - \nu \int_0^{2\pi}
    \sin ( \theta - \theta' + \alpha ) p^{\sigma'}(\theta',t) 
  d \theta'~.
\end{equation}

Equations \eref{eq:continuity} and \eref{eq:v_Def} constitute a partial integro-differential equation for the distribution of oscillators $p^{\sigma}(\theta,t)$ in each cluster.

\subsection{Ott-Antonsen reduction}
\label{sec:OttAntonsen}
In 2008, Edward Ott and Thomas Antonsen proposed a simplified approach to solving this system \cite{Ott2008}. They suggested expanding $p^{\sigma}$ in a Fourier series, and restricting analysis to a particular low-dimensional manifold defined by $a_n = a^n$, where $a_n$ is the $n$th Fourier coefficient. They subsequently showed that this manifold is globally attracting for a broad class of Kuramoto oscillators \cite{Ott2009_1,Ott2009_2} {such as those satisfying \eref{eq:kuramoto}} \footnote{{This manifold is only globally attracting for oscillators with non-singular (e.g. Gaussian, Lorentzian or sech) frequency distributions. For identical oscillators, the frequency distribution is a delta function, so the manifold is not globally attracting \cite{Ott2009_2}.  However, Pikovsky and Rosenblum demonstrated that for particular constants of motion \cite{Watanabe1994} the dynamics evolve along the Ott-Antonsen manifold \cite{Pikovsky2008}. Thus the Ott-Antonsen anzatz is useful for characterizing some of the dynamics even when oscillators are identical.}}. Paz\'{o} and Montbri\'{o} recently generalised this result by showing that Winfree oscillators {(see section \ref{sec:winfree})} also converge to the Ott-Antonsen manifold \cite{Pazo2014}.  

We therefore consider distributions with the form
\begin{equation}\label{eq:poisson}
%  2 \pi p^{\sigma}(\theta, t) = 1 + \sum^\infty_{n=1} \left( 
%    a_{\sigma}(t) e ^{i \theta} \right)^n + c.c.~,
  2 \pi p^{\sigma}(\theta, t) = 1 + \sum^\infty_{n=1} \left\{ 
    \left[ a_{\sigma}(t) e ^{i \theta} \right]^n + 
    \left[ a_{\sigma}^\ast(t) e ^{-i \theta} \right]^n 
    \right\}~.
\end{equation}
where the superscript $^\ast$ denotes complex conjugation. Substitution of equation \eref{eq:poisson} into equation \eref{eq:v_Def} reveals that
\begin{equation} \label{eq:v_Def2}
v^{\sigma} (\theta, t)=\omega -\frac{z_{\sigma}e^{i\alpha}}{2i} e^{i\theta}+ \frac{z_{\sigma}^*e^{-i\alpha}}{2i}e^{-i\theta}
%\omega - e^{i\theta}\frac{e^{i\alpha}}{2i}\left(\mu a_{\sigma}^*+\nu a_{\sigma^{\prime}}^*\right)+ e^{-i\theta}\frac{e^{-i\alpha}}{2i}\left(\mu a_{\sigma}+\nu a_{\sigma^{\prime}}\right)
%\omega - \mu \frac{a_{\sigma}^*e^{i(\theta+\alpha)}- a_{\sigma}e^{-i(\theta+\alpha)}}{2i}       - \nu \frac{a_{\sigma^{\prime}}^*e^{i(\theta+\alpha)}- a_{\sigma^{\prime}}e^{-i(\theta+\alpha)}}{2i}
\end{equation}
where we have defined 
%\begin{equation*}\label{eq:complexorder}
$  z_{\sigma}(t) = \mu \left< e^{i\theta_j^{\sigma}} \right>_{j\in \sigma} + 
         \nu \left< e^{i\theta_j^{\sigma^{\prime}}} \right>_{j\in \sigma^{\prime}} = 
         \mu a_{\sigma}^{\ast} + \nu a_{\sigma^{\prime}}^{\ast}~.
$ 
{Thus equation \eref{eq:continuity} becomes}
\begin{equation}  \label{eq:newcont}
\sum_{n=1}^\infty \left[
  c_n e^{i(n-1)\theta} + d_n e^{i n \theta} + f_n e^{i(n+1)\theta} + \textrm{c.c.}
  \right] = 
  \frac{1}{2} z^*_\sigma e^{i\alpha}e^{i\theta} + \textrm{c.c.}   ~,
\end{equation} {
where $c_n = \half (n-1) z_\sigma a_\sigma^n e^{-i \alpha}$, $d_n = n a_\sigma^{n-1} \dot{a}_\sigma + i n \omega a_\sigma^n$, and $f_n = -\half (n+1) z^*_\sigma a_\sigma^n e^{i \alpha}$.
Equating coefficients of $e^{i \theta}$ on the left and right-hand sides of \eref{eq:newcont} allows us to describe the dynamics of $a$ in each cluster as}
\begin{equation}\label{eq:amplitude}
  \frac{da_{\sigma}}{dt} + i \omega a_{\sigma} + \half \left[ 
    a_{\sigma}^2 z_{\sigma}e ^{-i\alpha} - z^*_{\sigma}e^{i \alpha}
  \right] = 0~.
\end{equation}

{\subsection{Simplified governing equations}}
Equation \eref{eq:amplitude} applies independently to each cluster.  For convenience we define $a_X = \rho_X e ^{- i \phi_{X}}$ and $ a_Y = \rho_Y e ^{- i \phi_{Y}}$ for clusters $X$ and $Y$, respectively, then use equation \eref{eq:amplitude} to find
\begin{eqnarray}\label{eq:polarampeqns}
  0 & = & \dot{\rho}_X + \frac{\rho_X^2 - 1}{2} \left[ \mu \rho_X
\cos \alpha + \nu \rho_Y \cos \left( \phi_Y - \phi_X - \alpha
\right) \right] \\
  0 & = &  -\rho_X\dot{\phi}_X  + \rho_X \omega - \frac{1+
\rho^2_X}{2} \left[ \mu \rho_X \sin \alpha + \nu \rho_Y \sin \left(
\phi_X - \phi_Y + \alpha\right) \right]~,  \nonumber
\end{eqnarray}
with analogous equations for $ \dot{\rho}_Y $ and $ \dot{\phi}_Y$.

Chimera states correspond to stationary solutions with $\rho_X=1$ and $\rho_Y < 1$ (and vice versa).  Fixing $\rho_X=1$, defining $r=\rho_Y$ and $\psi = \phi_X - \phi_Y$, we obtain the following system of equations for chimera states
\begin{eqnarray}\label{eq:2dsystem}
  \dot{r} & = & \frac{1-r^2}{2} \left[ \mu r \cos \alpha + \nu \cos
(\psi - \alpha) \right]  \\
  \dot{\psi} & = & \frac{1+r^2}{2r} \left[ 
    \mu r \sin \alpha - \nu \sin(\psi - \alpha) 
  \right] - \mu \sin \alpha - \nu r \sin ( \psi + \alpha )~. \nonumber
\end{eqnarray}
Solutions for and bifurcations of chimera states can now be found by analysis of the properties of this simple two-dimensional dynamical system.  An example of a chimera state in this system (with $r=0.729$ and $\psi=0.209$ and 1024 oscillators per cluster) is displayed in panel (a) of \fref{fig:examples}. 
\vspace{0.1pt}

%************ Section ************%
\section{What's known}

\subsection{Bifurcations of chimera states} \label{sec:bif}
Analysis of system \eref{eq:polarampeqns} reveals a chimera state ``life cycle'' as follows: When $\alpha = \pi/2$, both symmetric $\rho_X = \rho_Y$ states and asymmetric $\rho_X \neq \rho_Y$ states are possible.  In parallel with earlier work \cite{Abrams2006}, we refer to the symmetric states as ``uniform drift'' and the asymmetric states as ``modulated drift'' (where the descriptor indicates spatial uniformity or modulation---in both cases the drifting oscillators behave nonuniformly in time).   As $\alpha$ decreases from $\pi/2$, an unstable chimera bifurcates off of the fully synchronised state, while a stable chimera state bifurcates off the modulated drift state. Further decreasing $\alpha$ eventually results in a saddle-node bifurcation. \footnote{A third unstable chimera bifurcates off of the unstable anti-synchronised state ($r=1$, $\psi=\pi$) as $\alpha$ decreases from $\pi/2$ and it persists for all values of $\alpha$. Note that the stability results described above are only valid for $0<\alpha<\pi/2$.} 

When the coupling disparity $\mu-\nu$ becomes sufficiently large, chimera states can also undergo a Hopf bifurcation. This causes the order parameter for the incoherent cluster to oscillate, resulting in a `breathing' phenomenon. The order parameter $r e^{i \psi}$ follows a limit cycle in the complex plane, the diameter of which increases as $\mu-\nu$ increases.  At a critical value of $\mu-\nu$ that limit cycle collides with the unstable chimera state, resulting in the disappearance of the `breathing' chimera state through homoclinic bifurcation \cite{Abrams2008,Panaggio2014thesis}.

\begin{figure*}[bth!]
    \centering
	\includegraphics[width=0.9\textwidth]{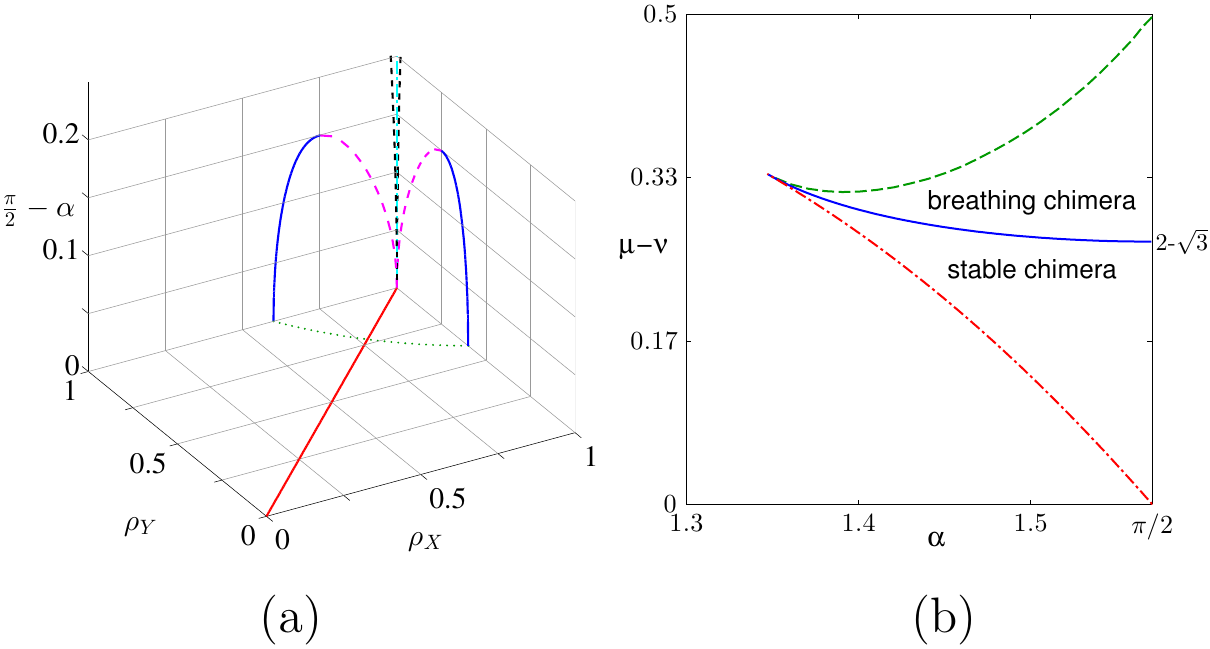}
	\caption{Two-cluster chimera. (a) Origin of chimera state via bifurcation off of modulated drift (green dotted) and uniform drift (red solid) states. Three chimera states are shown: stable (blue solid) and unstable (magenta dashed) chimeras, both with $\psi$ near $0$, and a second unstable chimera (black dashed) with $\psi$ near $\pi$. The stable fully synchronised state  with $\rho_X=\rho_Y=1$, $\psi=0$ and the unstable anti-synchronised state with $\rho_X=\rho_Y=1$, $\psi=\pi$ coincide in this projection and are indicated by the cyan dash-dotted line. Here equations \eref{eq:polarampeqns} are used with $\mu=0.625, \nu=0.375$. (b) Bifurcations of chimera state in parameter space of coupling strength disparity $\mu-\nu$ and phase lag $\alpha$ with $\mu+\nu=1$. Red dash-dotted line indicates saddle-node bifurcation, blue solid line indicates Hopf bifurcation, green dashed line indicates homoclinic bifurcation. Chimera states are detectable in between red dash-dotted line and green dashed line.} 
	\label{fig:bif_2_chimera}
\end{figure*}

These bifurcations are displayed in \fref{fig:bif_2_chimera}.  

\subsection{Chimeras on spatial networks}
Chimera states have been analyzed in a variety of different topological settings (see appendix), and the bifurcations described above appear to be generic. Thus far, chimeras {for traditional Kuramoto phase-oscillators (as described by equation \eref{eq:kuramoto})} have been reported on a ring of oscillators \cite{Kuramoto2002,Abrams2004,Abrams2006,Maistrenko2014}, {a finite strip with no-flux boundaries \cite{Zhu2012}}, two- and three-cluster networks \cite{Abrams2008,Martens2010}, and oscillators distributed along an infinite plane \cite{Kuramoto2003_2,Shima2004,Martens2009}, a torus \cite{Omelchenko2012,Panaggio2013} and a sphere \cite{Panaggio2014,Panaggio2014thesis}.  

{Depending on the topology, two distinct classes of chimera states may appear: spots and spirals.  In spot chimeras, synchronous oscillators all share nearly the same phase while incoherent oscillators have a distribution of phases.  When phase is indicated by color this creates a ``spot'' pattern where the coherent region is nearly monochromatic and the incoherent region contains specks of many different colors\footnote{This definition includes patterns with stripes as well as circular and irregularly shaped spots}.  Spots have only been reported for near-global coupling with $\alpha$ near $\pi/2$. The drifting and locked regions in these systems each occupy a finite fraction of the domain.  Spot chimeras occur in every system studied with the exception of the infinite plane.  On the plane, any finite-sized spot would represent an infinitesimal fraction of the domain, and as a result might be argued to be insignificant.  Spots and/or stripes with infinite size have not been reported at this time \footnote{{Recent investigations by Kawamura \cite{Kawamura2007} and Laing \cite{Laing2011} have revealed stripe chimeras that appear in arbitrarily large but finite networks. This strongly suggests that stripes with infinite size are also possible.}}.}

On two-dimensional surfaces, spiral chimeras can also occur.  These chimeras consist of an incoherent core surrounded by rotating spiral arms that are locally synchronised.  Along a path around the incoherent core, the phases of coherent oscillators make a full cycle.  Examples of these types of patterns can be found in  \fref{fig:examples}. Spiral chimeras have been reported on a plane \cite{Shima2004,Martens2010,Gu2013}, a torus (in configurations of 4 {or more} spirals) {\cite{Kim2004,Omelchenko2012}}, and on a sphere \cite{Panaggio2014,Panaggio2014thesis}. These spirals appear to be stable only when $\alpha$ is near $0$, and when the coupling kernel is more localised than for their spot counterparts.

\subsection{Chimeras on arbitrary networks}
Recently, the concept of a chimera state has been extended to networks without a clear spatial interpretation. Thus far, the evidence for chimera states on these networks is largely numerical.  Shanahan considered a network consisting of eight communities of 32 oscillators.  Oscillators were fully coupled to other oscillators within the same community and connected at random to 32 oscillators from the other communities.  He observed fluctuations in both internal and pairwise synchrony in the communities resembling chimera states \cite{Shanahan2010}.

Laing \etal analysed a two-cluster system with randomly removed links. They observed that chimera states are robust to small structural perturbations, but the ranges of parameter values for which they exist become increasingly narrow as the number of missing links increases \cite{Laing2012_2}.  Yao \etal performed a similar analysis of chimera states on a ring and confirmed that chimera states remain apparently stable after a small fraction of links have been removed \cite{Yao2013}.

Zhu \etal took a slightly different approach to this problem.  They considered randomly generated Erd\"{o}s-R\'{e}nyi and scale-free networks of identical oscillators.  In lieu of spatial structure, they classified oscillators using their effective angular velocities and found that certain oscillators became phase- and frequency-locked while other oscillators drifted. On scale-free networks, the highly connected hubs were more likely to synchronise than less connected oscillators.  On Erd\"{o}s-R\'{e}nyi networks, all oscillators seemed equally likely to remain coherent \cite{Zhu2014}.

\subsection{Stability of chimera states}
\label{sec:stabilityknown}
Rigorous analysis of the stability of chimera states has proven to be elusive.  In many papers, when chimera states are referred to as stable, the authors simply mean that they are states that persist in simulations with a finite duration and a finite number of oscillators. This heuristic approach can be useful for identifying unstable states, but it is unable to differentiate between stable states and long-lived transients.

The most successful analytical investigation of the stability of chimera states was carried out by Omel'chenko in 2013.  He examined a ring of oscillators described by equation \eref{eq:kuramoto} and showed that a variety of stationary ``coherence-incoherence'' patterns existed along the Ott-Antonsen manifold.  He then explored the stability of these solutions with respect to perturbations along this manifold by computing the point and essential spectra using the theory of compact operators.  He was able to demonstrate the existence of multiple pairs of stable and unstable solutions for arbitrary piecewise smooth, even and $2\pi-$periodic coupling functions $G(x)$. Thus, with an infinite number of oscillators, chimera states appear to be stable \cite{Omelchenko2013}.

For finite networks of oscillators, numerical experiments suggest that chimeras states on a ring are actually long-lived transients \cite{Wolfrum2011_2}.  To show this, Matthias Wolfrum and Oleh Omel'chenko considered a ring of oscillators with a finite coupling range $R$
\begin{equation}
\frac{d\psi_k(t)}{dt}=\omega-\frac{1}{2R}\sum_{j=k-R}^{k+R} \sin\left[\psi_k(t)-\psi_j(t)+\alpha\right]
\end{equation}
They computed the Lyapunov spectrum of the system and show that it corresponded to a `weakly hyper chaotic trajectory'; however, as the system size increased, the chaotic part of the spectrum tended to 0.  The lifetime of this transient trajectory grew exponentially with the system size \cite{Wolfrum2011_2}.  Omel'chenko \etal also found that the incoherent regions in these systems could drift when the number of oscillators was small, but that as the system grew, this finite size effect disappeared \cite{Omelchenko2010}. There is numerical evidence that these conclusions apply to other coupling schemes and non-identical frequencies, but this has not been shown conclusively \cite{Wolfrum2011_1,Wolfrum2011_2}.

Wolfrum and Omel'chenko along with Jan Sieber later showed that these chimera states could be stabilised by implementing a control scheme with time-dependent phase lag $\alpha(t)=\alpha_0+K(r(t)-r_0)$ where $\alpha_0$ and $r_0$ correspond to the desired final phase lag and global order parameter respectively \cite{Sieber2014}.

%************ Section ************%
\section{Generalizations}
Chimera states were first characterised on simple networks of identical Kuramoto-style phase oscillators. However, these patterns can also be observed in networks with more general types of oscillators.

\subsection{Non-constant amplitude}
In Kuramoto's original paper, he observed chimera states with both non-locally coupled Stuart-Landau oscillators\footnote{Note that there is some ambiguity in the literature regarding what is referred to as a Stuart-Landau oscillator and what is a Ginzburg-Landau oscillator.} (variable amplitude and phase) and with Kuramoto-style oscillators possessing a fixed amplitude.  It is straightforward to show that these systems are essentially the same when the coupling is weak. To see this, consider the Stuart-Landau equation (the complex Ginzburg-Landau equation without diffusion)  with a coupling term described by the operator $\mathcal{L} W(\xv,t)$:
\begin{eqnarray*}
\frac{\partial}{\partial t} W(\xv,t)&=&(1+ia)W(\xv,t)-(1+ib)W(\xv,t)\left|W(\xv,t)\right|^2 \nonumber \\& &+\epsilon e^{-i\alpha}\mathcal{L} W(\xv,t)~.
\end{eqnarray*}
where the vector $\xv$ indicates the location in a space of arbitrary dimension.  Let $W(\xv,t)=R(\xv,t)e^{i\theta(\xv,t)}$. After, dividing into real and imaginary parts and shifting into a rotating frame of reference $\phi(\xv,t)=\theta(\xv,t)-(a-b)t$, we find that to leading order in $\epsilon$
\begin{eqnarray}
\frac{\partial}{\partial t} R(\xv,t)& = & R(\xv,t)-R(\xv,t)^3 +O(\epsilon)\\
\frac{\partial}{\partial t} \phi(\xv,t)& = & b(1-R(\xv,t)^2)+O(\epsilon)~.
\end{eqnarray}
Thus, there is a separation of time scales when $\epsilon$ is small.  On the fast time scale, oscillators approach a stable limit cycle with amplitude $R(\xv,t)\approx 1$ where deviations are order $\epsilon$ or smaller. After fixing $R(\xv,t)=1$, on the slow time scale, the dynamics can be expressed in terms of $\phi(\xv,t)$.  For the particular case of non-local coupling $ \mathcal{L} W(\xv,t) = \int_S G(\xv-\xvp)W(\xvp,t)d\xvp-W(\xv,t)$, where $G(\xv)$ represents a coupling kernel, the phase equation becomes (to lowest order)  
%\begin{equation} 
%\frac{\partial}{\partial t} \phi(\xv,t)=\epsilon\left(\sin\alpha+\cos{\alpha}\int_S G(\xv-\xvp)\sin(\phi(\xv,t)-\phi(\xvp,t))d\xvp-\sin{alpha}\int_S G(\xv-\xvp)\cos(\phi(\xv,t)-\phi(\xvp,t))d\xvp\right).
%\end{equation}
\begin{equation}
\frac{\partial}{\partial t} \phi(\xv,t)=\omega-\epsilon\int_S G(\xv-\xvp)\sin(\phi(\xv,t)-\phi(\xvp,t)+\alpha)d\xvp
\end{equation}
where $\omega=\epsilon\sin\alpha$. This is the continuum Kuramoto model. Discretising the domain and defining $K_{ij}=G(\xv_i-\xv_j)$, we obtain the more familiar discrete formulation
\begin{equation}
\label{discrete_kuramoto}
  \frac{\partial}{\partial t} \phi_i(t) = \omega-\frac{\epsilon}{N} \sum_{j=1}^NK_{ij} \sin(\phi_i(t)-\phi_j(t)+\alpha)~.
\end{equation}

Most of the literature on chimera states deals with Kuramoto oscillators, however, it appears that coupled Stuart-Landau oscillators behave similarly \cite{Sethia2013}.  For example, Carlo Laing considers a generalisation of the two-cluster chimera for Stuart-Landau oscillators. He shows that the expected bifurcations persist even when the amplitude of oscillation is allowed to vary \cite{Laing2010}.

Kuramoto and Shima demonstrated that spiral chimeras can also be sustained by Stuart-Landau oscillators on a plane. They considered the standard non-locally coupled complex Ginzburg-Landau equation and reported that with a coupling kernel $G(x) \propto K_0(x/\sqrt{D})$ (where $K_0$ is a modified Bessel function of the second kind) it was possible to observe spiral waves surrounding an incoherent core \cite{Kuramoto2003_2}.

The additional degree of freedom for Stuart-Landau oscillators can allow for more complex dynamics as well. Bordyugov, Pikovsky and Rosenblum considered a ring of oscillators with length $2\ell$ governed by the equation
\begin{equation}
\frac{\partial A}{\partial t}=(1+i\omega)A-\left|A\right|^2A+\epsilon Z
\end{equation}
where $Z = B e^{i\beta_0}e^{i\beta_1\left|B\right|^2}$, $B = \int_{-\ell}^{\ell}G(x-x^{\prime})A(x^{\prime},t)dx^{\prime}$ and $G=ce^{-\left|x\right|}$. This represents non-local coupling {with phase lag that varies in space (and with amplitude)}. The authors explored the role of the coupling distance relative to the system size and observed a parameter regime where the synchronised state was unstable and where chimera states appeared spontaneously. In addition to traditional chimera states, the authors reported the existence of `turbulent chimeras' in which regions of local synchronisation appeared and vanished {seemingly} randomly over time \cite{Bordyugov2010}.

Zakharova \etal studied asymmetrically-coupled Stuart-Landau oscillators and demonstrated that increases in the coupling range can lead to chimera death, a phenomenon in which a chimera state breaks down and all oscillation ceases \cite{Zakharova2014}.

\subsection{Winfree model} \label{sec:winfree}
The Kuramoto model can also be derived as a special case of the Winfree model {\cite{Winfree1967,Winfree2001}}.  To see this, consider the Winfree model with a pulse shape $P(\theta)$ and response curve $Q(\theta)$
\begin{equation}
  \frac{d}{dt} \theta_i = \omega_i + \frac{\epsilon}{N} Q(\theta_i) \sum_{j=1}^N P(\theta_j)~.
  \label{eq:winfree}
\end{equation}

When the coupling is sufficiently weak and the oscillators are nearly identical, the phase can be replaced by its average over an entire period, yielding
\begin{equation}
  \frac{d}{dt} \theta^{\textnormal{avg}}_i = \omega_i+\frac{\epsilon}{N}\sum_{j=1}^N \frac{1}{2\pi} \int_{-\pi}^{\pi} Q(\theta^{\textnormal{avg}}_i + \lambda) P(\theta^{\textnormal{avg}}_j + \lambda)d\lambda~.
\end{equation}
The integral can be evaluated for a variety of smooth functions $P$ and $Q$; it is especially simple for sinusoidal $Q$ and peaked $P$.  As an example, take $Q(\theta)=-\sin(\theta+\alpha)$ and $P(\theta)=2 \pi \delta(\theta)$. By the sifting property of the Dirac delta function,  
\begin{equation*}
  -\int_{-\pi}^{\pi} \sin(\theta^{\textnormal{avg}}_i + \lambda + \alpha) \delta(\theta^{\textnormal{avg}}_j + \lambda) d\lambda = -\sin(\theta^{\textnormal{avg}}_i - \theta^{\textnormal{avg}}_j + \alpha)~,
\end{equation*}
and thus the Winfree model simplifies to
\begin{equation}
  \frac{d}{dt} \theta^{\textnormal{avg}}_i = \omega_i-\frac{\epsilon}{N}\sum_{j=1}^N\sin(\theta^{\textnormal{avg}}_i - \theta^{\textnormal{avg}}_j + \alpha)~,
\end{equation}
which is just the familiar Kuramoto model.  The Kuramoto model can also be derived for a variety of smooth finite pulse functions $P(\theta)$. 

In a 2014 publication in Physical Review X, Paz\'{o} and Montbri\'{o} demonstrated that Winfree oscillators also have solutions on the invariant manifold proposed by Ott and Antonsen \cite{Ott2008}. This allows for a reduction to a system of three ordinary differential equations for a two-cluster network and two integro-differential equations for networks with non-local coupling. For Kuramoto oscillators, this development opened up the possibility of analytically characterising chimera states.  It remains to be seen whether many of the subsequent results for chimera states can be generalised to Winfree oscillators \cite{Pazo2014}.

\subsection{Nonidentical oscillators}
Although symmetry breaking phenomena like chimera states are particularly surprising when the oscillators are identical, these patterns are certainly not unique to identical oscillators. In 2004, Montbri\'o \etal reported the coexistence of coherence and incoherence in the two-cluster network of oscillators with a Lorentzian frequency distribution. Unlike the homogeneous case, coexistence was possible for all values of $\alpha$ \cite{Montbrio2004}. Later, Carlo Laing performed extensive analysis on the two-cluster network, one-dimensional ring, and infinite plane and showed that key results pertaining to chimera states in those systems could be generalised to oscillators with heterogeneous frequencies \cite{Laing2009_1,Laing2009_2}.  He demonstrated that these heterogeneities can lead to new bifurcations allowing for alternating synchrony between the distinct populations over time. He also showed that chimera states are robust to temporal noise \cite{Laing2012}.

\subsection{Inertial oscillators}
Chimera states are possible in systems with inertia as well. Bountis \etal studied a variation on the two-cluster network with non-identical phase oscillators, motivated by equations for coupled pendula. They found that chimera states continued to appear in simulation as long as the inertial terms were small. In addition, they observed that chimera states ceased to exist when the magnitude of the first derivative term (representing dissipation) dropped below a critical threshold \cite{Bountis2014}.

\subsection{Return maps}
Chimera states occur in another type of oscillatory system: iterated maps. Iryna Omelchenko \etal showed that a ring of coupled chaotic maps can exhibit chimera-like phenomena \cite{Omelchenko2011}.  They considered the system
\begin{equation}
z_i^{t+1}=f(z_i^t)+\frac{\sigma}{2P}\sum_{j=i-P}^{i+P}\left[f(z_j^t)-f(z_i^t)\right]
\end{equation}
where $z_i^t$ is analogous to the phase of oscillator $i$ at step $t$ and $f$ is the logistic map $f(z)=3.8z(1-z)$.  Depending on the coupling distance $P$ and coupling strength $\sigma$, they observed fixed points consisting of regions of synchrony separated by narrow bands of incoherence.

%************ Section ************%
\section{Experiments} \label{sec:experiments}

For an entire decade, chimera states were observed only in numerical simulations. Many of these chimeras required carefully chosen initial conditions and seemed to be sensitive to perturbations.  So, it was unclear whether chimera states were robust enough to be observed in experiments.  

Then in July 2012, this question was answered definitively when two successful experimental chimeras, one at West Virginia University and the other at the University of Maryland, were reported in Nature Physics \cite{Tinsley2012,Hagerstrom2012}. The first group, led by Kenneth Showalter, used the Belousov-Zhabotinsky reaction to create a realisation of a two-cluster chimera similar to the one reported in ref.~\cite{Abrams2008}.  They divided a population of photosensitive chemical oscillators into two separate groups and used light to provide feedback for the reactions.  Oscillators were weakly coupled to the mean intensity of the oscillators within the opposite group and more strongly coupled to the intensity of other oscillators within the same group with a fixed time-delay. They observed a variety of dynamical patterns including complete synchronisation, synchronised clusters and chimera states in which only one of the two groups synchronised \cite{Tinsley2012}. They later carried out a similar experiment on a non-locally coupled one-dimensional ring of oscillators and observed a variety of chimera-like patterns resembling those seen in theoretical studies \cite{Nkomo2013}.

Simultaneously, Thomas E. Murphy, Rajarshi Roy, and graduate student Aaron M. Hagerstrom designed a coupled map lattice consisting of a spatial light modulator controlled by a computer with feedback from a camera.  This was essentially a realisation of the chaotic maps studied by Omelchenko \etal\cite{Omelchenko2011}.  Roy's group reported chimeras on both one-dimensional rings and two-dimensional lattices with periodic boundaries \cite{Hagerstrom2012}. 

One critique of these experiments was their reliance on computers to provide coupling between the oscillators and maps \cite{Smart2012}. However, these concerns were addressed by a third experiment that relied on mechanical coupling alone.  Erik Martens and his colleagues placed metronomes on swings coupled by springs.  The vibrations of the swings provided strong coupling between oscillators on the same swing, and the springs weakly coupled metronomes on opposite swings.  By varying the spring constant they were able to observe chimera states along with the expected in-phase and anti-phase synchronous states \cite{Martens2013}.

More recently, a group in Germany has observed chimera states that form spontaneously in a photoelectrochemical experiment. They model the oxidation of silicon using a complex Ginzburg-Landau equation with diffusive coupling and nonlinear global coupling. Schmidt \etal report that in numerical simulations and experiments, the thickness of an oxide layer exhibits coexisting regions of synchronous and asynchronous oscillation \cite{Schmidt2014}.

%************ Section ************%
\section{Possible applications}

Chimera states have not been conclusively determined to exist outside of laboratory settings, but there are many natural phenomena that bear a strong resemblance to chimera states and may be linked to these types of dynamics.

\subsection{Unihemispheric sleep} \label{sec:sleep}
Many species including various types of mammals and birds engage in unihemispheric slow-wave sleep.  In essence, this means that one brain hemisphere appears to be inactive while the other remains active. The neural activity observed in EEGs during this state reveals high-amplitude and low frequency electrical activity in the sleeping hemisphere, while the other hemisphere is more erratic \cite{Rattenborg2000}. The chimera states observed in ref.~\cite{Abrams2008} can be interpreted as a model of coordinated oscillation in one hemisphere and incoherent behaviour in the other. Typically, these activity patterns alternate between hemispheres over time. Ma, Wang and Liu attempted to reproduce this alternating synchronisation. They considered the model
\begin{equation}\label{eq:gov2}
  \frac{d \theta^\sigma_i}{dt} = \omega_i +  
    \sum^2_{\sigma^{\prime} = 1} \frac{K_{\sigma \sigma^{\prime }}}{N_{\sigma^{\prime}}}
    \sum^{N_{\sigma^{\prime}}}_{j=1} \sin ( \theta^{\sigma^{\prime}}_j 
    - \theta^\sigma_i - \alpha )+A\sin\Omega(t-\tau_{\sigma})
\end{equation}
and found that if $\tau_1\neq \tau_2$ (different reactions to environmental forcing), for appropriate choices of coupling strengths periods of coherence and incoherence alternated in each hemisphere \cite{Ma2010}.

\subsection{Ventricular fibrillation}
\begin{figure*}[bth!]
    \centering
	\includegraphics[width=0.8\textwidth]{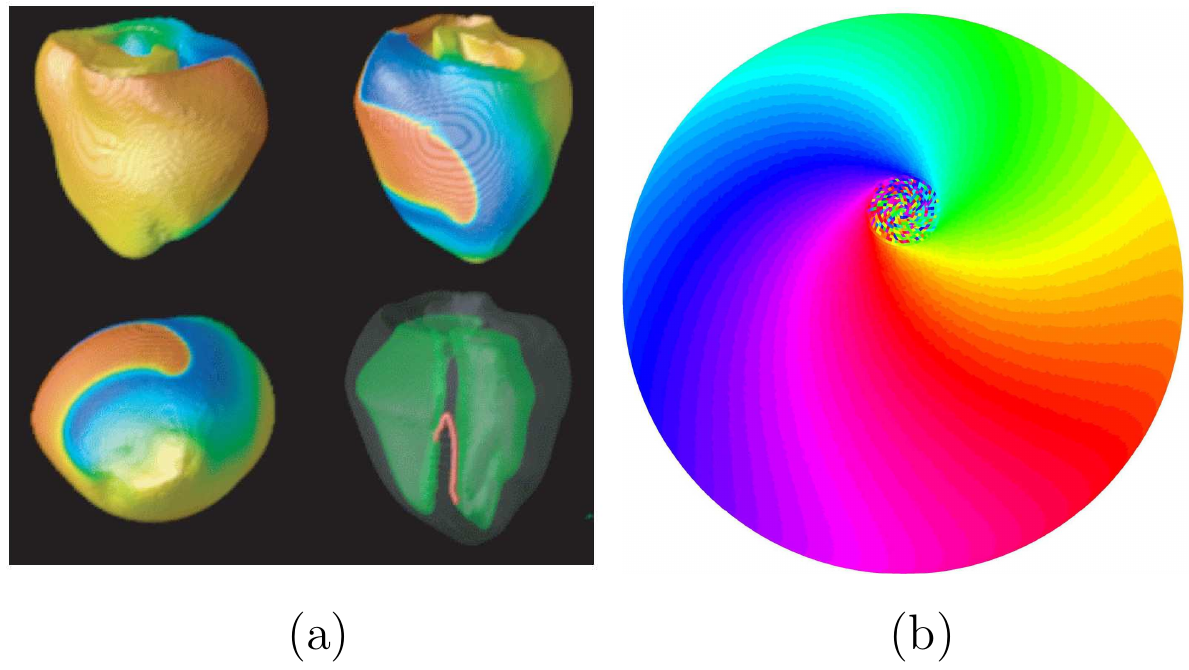}
	\caption{Spiral Waves. (a) A spiral wave on the surface of a human heart. Reproduced with permission from \cite{Cherry2008}. \copyright~IOP Publishing \& Deutsche Physikalische Gesellschaft. CC BY-NC-SA. (b) A spiral wave chimera state on the surface of a sphere.} 
	\label{fig:spirals}
\end{figure*}

Ventricular fibrillation is one of the primary causes of sudden cardiac death in humans. This phenomenon results from a loss of coordination in the contractions of cells within the heart.  During fibrillation, spiral wave patterns can form \cite{Davidenko1992, Panfilov1998, Winfree1983}.  At the center of these rotating patterns, there is a phase singularity and the dynamics are unclear.  The contractions near this singularity may be uncoordinated.  These types of patterns are also observed in coupled oscillators arranged on the surface of a sphere.  In these arrays, when the phase lag is non-zero, a finite fraction of oscillators at the center of the spiral wave remain incoherent.  Thus, spiral wave chimeras may be viewed as a model for the patterns formed by the contractions of heart cells during ventricular fibrillation (figure \ref{fig:spirals}).  

\subsection{Power grid}
The U.S.~power grid consists of many generators producing power at a frequency of about 60 Hz. Under ideal conditions, the generators are synchronised.  Synchronisation of a power grid is often studied using Kuramoto-like models {(e.g., \cite{Filatrella2008,Dorfler2012,Dorfler2013,Motter2013})}. Analysis of these models has shown that a variety of perturbations to the network can cause full or partial desynchronisation, which may lead to blackouts.  Knowledge of the possibility of chimera states in power distribution networks---and the chimera state basins of attraction---could be useful for maintaining stable and robust synchrony.

\subsection{Social systems}
Chimera-like states may also be possible in social systems.  Gonz\'{a}lez-Avella \etal examined a model for the dissemination of social and cultural trends. They observe that coupled populations can exhibit chimera-like patterns in which consensus forms in one population while the second population remains disordered\cite{GonzalezAvella2014}.

\subsection{Neural systems}
{Chimera states bear a strong resemblance to bump states observed in neural networks---localized regions of coherent oscillation surrounded by incoherence (see, e.g.,  \cite{Compte2000, Laing2001, Laing2009_1, Shanahan2010, Wimmer2014}).  In certain models, they appear to form when fronts between regions of coherence and incoherence
collide \cite{Laing2011}. Bumps appear to be stable in networks of delay-coupled Kuramoto oscillators and in more complex models of neural oscillators. } For example, Laing and Chow studied networks of integrate-and-fire neurons, a type of pulse-coupled oscillator. They observed solutions with a spatially dependent firing rate.  Outside of the bump oscillators do not fire and inside they fire asynchronously \cite{Laing2001}.  {Chimera-like patterns} have also been reported for non-locally coupled Hodgkin-Huxley oscillators \cite{Sakaguchi2006}, Fitz-Hugh Nagumo oscillators \cite{Omelchenko2013_2}, leaky integrate-and-fire neurons \cite{Olmi2010}, in the lighthouse model \cite{Chow2006}, and in many other neural network models \cite{Compte2000,Renart2003}.

There is observational evidence of chimera-like states in electrical brain activity. Tognoli and Kelso report that, during studies where participants were asked to coordinate left and right finger movement with a periodically flashing light, EEGs reveal clusters of coordinated and uncoordinated activity \cite{Tognoli2014}.

%************ Section ************%
\section{Open questions}

Over the last 12 years many significant advances in our understanding of chimera states have been made. Nonetheless, some important questions have yet to be answered conclusively.

\subsection{How does the phase-lag affect the dynamics?}
The Kuramoto model is often written in terms of a coupling phase lag parameter $\alpha$:
\begin{equation}
\frac{\partial}{\partial t} \phi_i(t)=\omega-\frac{\epsilon}{N}\sum_{j=1}^NK_{ij} \sin(\phi_i(t)-\phi_j(t)+\alpha)~.
\end{equation}
There are two natural interpretations for this parameter.  First, the phase lag can be interpreted as an approximation for a time-delayed coupling when the delay is small \cite{Crook1997}.  To see this, consider the system
\begin{equation}
\frac{\partial}{\partial t} \phi_i(t)=\omega-\frac{\epsilon}{N}\sum_{j=1}^NK_{ij} \sin(\phi_i(t)-\phi_j(t-\tau))~.
\end{equation}
When $\tau \ll 2 \pi / \omega$ and $\epsilon$ sufficiently small, 
\begin{eqnarray*}
  \phi_j(t-\tau) & \approx &\phi_j(t) - \tau\frac{d\phi_j(t)}{dt}
    \approx \phi_j(t) - \tau(\omega+O(\epsilon))\\
  & \approx &\phi_j(t) - \alpha \textnormal{ where } \alpha=\tau \omega.
\end{eqnarray*}
Thus phase lag can be thought of as a proxy for time delay that allows us to replace a system of an effectively infinite-dimensional delay differential equations with a system of ordinary differential equations.  Further examination of this point (as well as the loss of generality inherent in sinusoidal coupling) is found in reference \cite{Crook1997}.

A second interpretation can be seen by observing that the coupling term can be rewritten as 
\begin{eqnarray*}
  \sum_{j=1}^NK_{ij} \sin(\phi_i-\phi_j+\alpha) &=& \cos(\alpha)\sum_{j=1}^NK_{ij} \sin(\phi_i-\phi_j)  \\& & + \sin(\alpha)\sum_{j=1}^NK_{ij} \cos(\phi_i-\phi_j)~.
\end{eqnarray*}
When $\alpha=0$, only the sine coupling remains. In this case, complete synchronisation is the norm.  When $\alpha=\pi/2$, pure cosine coupling results in an integrable Hamiltonian system \cite{Watanabe1993,Watanabe1994}; this causes disordered initial states to remain disordered. Thus $\alpha$ determines a balance between spontaneous order and permanent disorder.

As mentioned previously, spiral and spot chimeras appear in different regions of parameter space. Stable spirals have been observed only when $\alpha$ is near 0 whereas spots only appear when $\alpha$ is near $\pi/2$. Thus spots occur near the Hamiltonian limit and spirals appear near the maximally dissipative limit \footnote{Perturbations off of the fully synchronised state can be shown to decay most rapidly when $\alpha=0$.}.  This observation has yet to be explained from an analytical perspective.

\subsection{What new dynamics appear when delay coupling is introduced?}
The Kuramoto model represents an idealisation of the interactions between coupled oscillators that might occur in natural systems.  However, a more realistic model for these interactions might incorporate time-delays in addition to or instead of a phase lag. {Adding time-delay into a model drastically increases the dimensionality of a system  making analysis more challenging. These additional degrees of freedom allow for more complex dynamics enabling even single oscillators to exhibit intervals of coherent and incoherent oscillation \cite{Larger2013}. For example,} Ma \etal considered a two-cluster network with uniformly distributed time-delays and phase lag.  They demonstrated that chimera states were robust to small delays.  They also showed that periodic forcing of the system can induce a chimera state in which the two clusters alternate between coherence and incoherence out of phase with each other.  This bears a resemblance to the patterns of brain activity during uni-hemispheric sleep \cite{Ma2010} (see also section \ref{sec:sleep} above).

Sethia, Sen and Atay examined the case of distance dependent delays on a ring of oscillators.  They showed that this type of coupling allows for `clustered' chimera states in which multiple regions of coherence are separated by narrow bands of incoherence \cite{Sethia2008}.

Another type of chimera state was reported by Sheeba \etal. They studied a two-cluster network with time-delay and reported that in addition to the traditional chimera states, one can also observe `globally clustered' chimera states in which the coherent and incoherent regions span both clusters \cite{Sheeba2009,Sheeba2010}.

\subsection{What are the necessary conditions for a chimera state?}
For years it was hypothesised that non-local/non-global coupling and phase-lag or time-delay were necessary for a chimera state to appear. However, recent results appear to contradict this hypothesis.

Omel'chenko \etal considered a system with global coupling and `spatially modulated' time-delayed coupling and non-periodic boundaries.  They showed that the spatial dependence in the strength of the delay coupling is sufficient to induce both stable and unstable chimera states that bifurcate from the coherent and incoherent states respectively and are destroyed in a saddle node bifurcation \cite{Omelchenko2008}.

Ko and Ermentrout showed that chimera-like states were also possible when the coupling strengths were heterogeneous.  They considered a network of Kuramoto oscillators with global coupling and non-zero phase lag, but with coupling strengths that followed a truncated power-law distribution.  They observed that, counter-intuitively, oscillators with weak coupling tended to synchronise while strongly coupled oscillators remained incoherent \cite{Ko2008}.

Wang and Li examined a system with global coupling that was weighted by the frequencies of heterogeneous oscillators. This allowed for both positive and negative coupling.  In their model, oscillators with negative natural frequencies remained incoherent while oscillators with positive frequencies synchronised \cite{Wang2011}.

Schmidt \etal studied an ensemble of Stuart-Landau oscillators with nonlinear mean-field coupling. They found that oscillators spontaneously split into a coherent cluster, in which all oscillators have the same amplitude and oscillate harmonically, and an incoherent cluster, in which amplitudes and phases are uncorrelated. They also showed that similar results could be observed in an experiment with electrochemical oscillators (see section \ref{sec:experiments}) \cite{Schmidt2014}. 

Sethia and Sen showed that mean-field coupling need not be nonlinear to allow for chimera states. They investigated a system of Stuart-Landau oscillators coupled through the mean field and found that oscillators can split into two groups: one exhibiting coherent oscillation and another with incoherent oscillation of smaller amplitude \cite{Sethia2014}.

Sethia and Sen also pointed out that Kuramoto had seen this behaviour in simulation years earlier. In a 1993 paper with Nakagawa, long before chimera states in phase oscillators were discovered, they observed synchronised and desynchronised clusters in globally coupled Stuart-Landau oscillators.  Although they did not delve into this phenomenon any further, Nakagawa and Kuramoto did make the astute observation that with global coupling ``the phase diagram is extremely simple in the weak coupling limit [phase oscillators]; the oscillators are either perfectly synchronised or completely independent \ldots No complex behaviour such as clustering and chaos can occur\ldots However, the origin of clustering discussed below is different because it comes from amplitude effects [71].''  It was not until 2014 that those findings were connected to chimera states \cite{Sethia2014}.

These results suggest that non-local/non-global coupling is not necessary for a chimera state to appear. Instead, non-uniformity may be all that is needed. This can be induced through variable coupling strength,  non-constant phase lag or time-delay, and by allowing for variation in the amplitude of oscillation.

\subsection{When are chimera states stable?}
\label{sec:stabilityquestion}
As mentioned in Section \ref{sec:stabilityknown}, chimera states on a ring have been shown to be stable in the limit $N \to \infty$ \cite{Omelchenko2013}. For finite $N$, where the Ott-Antonsen approach (see Section \ref{sec:OttAntonsen}) is not immediately applicable, no analytical stability results yet exist.  However, strong numerical evidence suggests that the chimera state on a ring is an extremely long lived transient for $N < \infty$ \cite{Wolfrum2011_2}.

It is unknown to what degree these results can be generalized to other networks of oscillators. Recent analyses of two-cluster (as in Section \ref{sec:simpleexample}) and multi-cluster systems suggest that chimera states are stable with as few as two oscillators per cluster \cite{Ashwin2014,Panaggio2015}. The discrepancy between the stability of chimera states on a one dimensional ring and chimera states on an effectively zero dimensional two-cluster network may result from the underlying network configuration. 

In higher dimensional (1D and above) spatially embedded networks with a finite number of oscillators (like a ring), the boundaries between incoherence and coherence typically move erratically throughout the space.  The drifting synchronization boundaries allowed in these systems may lead to an instability that is absent in clustered systems where drifting is precluded due to the fact that the coupling structure determines the boundaries on the incoherent and coherent regions. However, this hypothesis has yet to be confirmed analytically.  In arbitrary but finite networks without a clear spatial interpretation, it is unclear whether chimera states, if they exist, would be stable or transient or even whether rigorous and general stability analysis is possible.

%As mentioned in Section \ref{sec:stabilityknown}, chimera states on a ring appear to be chaotic transients, only gaining stability in the $N \to \infty$ limit. However, much remains to be explained: is rigorous stability analysis possible for finite $N$, where the Ott-Antonsen approach (see Section \ref{sec:OttAntonsen}) is not immediately applicable?  Recent results \cite{Ashwin2014,Panaggio2015} suggest that, in contrast to the behavior on the ring, chimera states in two-cluster (as in Section \ref{sec:simpleexample}) or multi-cluster systems are stable for finite $N$, perhaps even for $N$ as small as four.

%How do these stability results depend on the geometry of the underlying configuration space?  Irregular motion of chimera states in space is typically observed for finite $N$ in all systems where such motion is possible (1D or higher dimensional)---perhaps the dimensional impossibility of such ``drift'' in multi-cluster systems allows for finite-N stability?  If so, how would this generalize to arbitrary networks of oscillators where dimension of the configuration space is unclear?

\subsection{Is the existence of chimera states related to resonance?} \label{sec:resonance}
In their 2013 experiment involving two groups of metronomes on coupled swings, Martens \etal observed in-phase and anti-phase coherent solutions in which the oscillators on each swing synchronised and each swing behaved as a single pendulum.  These solutions occurred in different regions of phase space separated by a band of chimera states.  This band of chimeras was centred around the resonance curve for the anti-phase eigenmode. Martens \etal theorized that chimera states resulted from competition between the in-phase and anti-phase states and that they were a type of resonance phenomenon \cite{Martens2013}.  It is unclear if this observation is due to the fact that their model includes inertia, which is ignored in most phase-oscillator models, or whether this result can be generalized.

In another intriguing paper, Kawamura considered a system of non-locally coupled oscillators arranged along an infinite one-dimensional domain with parametric forcing \cite{Kawamura2007}. He noticed that when the forcing frequency was nearly twice the natural frequency it was possible for the  oscillators in the left and right halves of the domain to synchronise locally while remaining out of phase with oscillators in the other half. This resulted in a phase discontinuity at the origin.  For some parameter values, this discontinuity turned into a region of incoherence, producing a chimera state.  The fact that this occurred at twice the natural frequency suggests that this result may also be related to resonance.

\subsection{For what types of networks can chimera states exist?} \label{sec:nets}
The goal of making sense of the various incarnations of chimera states goes beyond just deepening our understanding of this still-puzzling phenomenon. Recently, Nicosia \etal found an intriguing connection between network symmetries and partially synchronised states for coupled oscillators \cite{Nicosia2013}. All numerical simulations that show chimera states are in fact represented in the computer as finite networks of some sort. If the theory for chimera states can be extended to more general networks, the range of applicability will be greatly enhanced---perhaps chimera state analogs exist on, e.g., the power grid, gene regulatory networks, and food webs? Maybe these states have been seen, either in the real world or in simulation, but have not been recognized or understood? If successful, a generalized theory connecting chimera states to topology and ultimately network structure would be a valuable tool.

%************ Section ************%
\section{Conclusion} \label{sec:conclusion} 
Given that oscillation is a nearly universal dynamical behaviour for physical systems, it is of fundamental interest to know just what can happen when oscillators are coupled together. Kuramoto's pioneering work in 2002 demonstrated that even networks of identical oscillators can have unexpected and counter-intuitive dynamics. These chimera states went unnoticed for decades due to their bistability with the spatially uniform states, but they have now been seen in a diverse set of analyses, numerical simulations and experiments. The robustness of these states and the diversity of the systems that are known to support them suggest that these patterns may occur naturally in some physical systems. Should chimera states be found outside of laboratory settings, identifying the types of interactions that can promote these behaviours could have profound practical implications.

\ack
The authors thank Erik Martens and Carlo Laing for useful feedback and discussions. They also thank the referees along with Ernest Montbri\'{o}, Tae-Wook Ko, Emmanuelle Tognoli, Gautam Sethia, and Laurent Larger for pointing out omissions in an early draft of this manuscript.

%************ Appendix ************%
\appendix
\section{Rough table of systems explored}

%\begin{landscape}
\begin{center}
%\begin{longtable}{@{}p{3cm}p{7cm}p{7cm}p{1.65cm}}
\begin{longtable}{@{}p{2cm}p{3.7cm}p{4.5cm}p{1.65cm}}
\caption[Feasible triples for a highly variable Grid]{Rough summary of systems and coupling functions explored. Note that, for compactness, notation is sometimes not identical to reference(s).  $x$ represents distance between spatial positions in 1D, $r$ represents distance between spatial positions in 2D, $d_{ij}$ represents shortest-path distance between nodes in a network, $(u,v)$ represent coordinates in 2D periodic space (torus). If not specified, oscillators studied are Kuramoto phase oscillators. Two-cluster systems have stronger intra-cluster and weaker inter-cluster coupling unless otherwise specified.  ``Top-hat'' coupling means constant coupling strength to all oscillators within some distance and zero coupling to oscillators beyond that distance.} \label{table:pastwork} \\

\br %\hline 
\multicolumn{1}{l}{Geometry} & 
\multicolumn{1}{l}{Coupling} & 
\multicolumn{1}{l}{Comments} & 
\multicolumn{1}{l}{Ref.(s)} \\ 
\mr %\hline 
\endfirsthead

\multicolumn{4}{c}%
{{\bfseries \tablename\ \thetable{} -- continued from previous page}} \\
\br %\hline 
\multicolumn{1}{l}{Geometry} &
\multicolumn{1}{l}{Coupling} &
\multicolumn{1}{l}{Comments} &
\multicolumn{1}{l}{Ref.(s)} \\ 
\mr %\hline 
\endhead

\mr %\hline 
\multicolumn{4}{r}{{Continued on next page}} \\ % \hline
\endfoot

\br %\hline \hline
\endlastfoot

0D 1-osc.
  & Time-delay
  & Virtual chimeras in fast time, FM electronics experiment
  & \cite{Larger2013} \\
0D 1-cluster and 2D plane
  & Nonlinear mean-field
  & Stuart-Landau, Ginzburg-Landau and experiment \\
0D 1-cluster
  & 
  & Scale-free dist.~of coupling strengths
  & \cite{Ko2008} \\
0D 1-cluster
  & 
  & Frequency-weighted coupling strengths, heterogeneous frequencies
  & \cite{Wang2011} \\
0D 1-cluster
  & Mean-field
  & Stuart-Landau oscillators
  & \cite{Sethia2014} \\
0D 2-cluster
  & 
  & Solvable
  & \cite{Abrams2008} \\
0D 2-cluster
  & 
  & Heterogeneous frequencies
  & \cite{Montbrio2004} \\
0D 2-cluster
  & 
  & Winfree (pulse-coupled) oscillators
  & \cite{Pazo2014} \\
0D 2-cluster
  & 
  & Stuart-Landau oscillators
  & \cite{Laing2010} \\
0D 2-cluster
  & 
  & Random connections
  & \cite{Laing2012_2} \\
0D 2-cluster
  & 
  & Heterogeneous frequencies, noise
  & \cite{Laing2012} \\
0D 2-cluster
  & 
  & Inertia
  & \cite{Bountis2014} \\
0D 2-cluster
  & 
  & Experiment (chemical oscillators), delay
  & \cite{Tinsley2012} \\
0D 2-cluster
  & 
  & Experiment (mechanical oscillators), inertia
  & \cite{Martens2013} \\
0D 2-cluster
  &  
  & Delay, forcing, asymmetry
  & \cite{Ma2010,Sheeba2009,Sheeba2010} \\
0D 2-cluster
  &  
  & Agent-based model, random
  & \cite{GonzalezAvella2014} \\
0D 2-cluster
  &  
  & LIF neurons
  & \cite{Olmi2010} \\
0D 3-cluster
  & 
  & Triangle $\to$ chain
  & \cite{Martens2010} \\
0D 8-cluster
  & Stronger intra-, weaker inter-cluster
  & Random inter-cluster connections
  & \cite{Shanahan2010} \\
0D multi-cluster
  & Arbitrary 
  & Examined validity of Ott-Antonsen ansatz \cite{Ott2008}
  & \cite{Pikovsky2008} \\
0D and 1D
  & 2-cluster and $G(x) \propto 1+A\cos(x)$
  & Heterogeneous frequencies
  & \cite{Laing2009_1} \\
1D periodic 
  & $G(x) \propto \exp(-\kappa|x|)$ 
  & First report of chimera state 
  & \cite{Kuramoto2002} \\
1D periodic 
  & $G(x) \propto \exp(-\kappa|x|)$ 
  & Delay from signal prop.
  & \cite{Sethia2008} \\
1D periodic 
  & $G(x) \propto 1+A\cos(x)$ 
  & 
  & \cite{Abrams2004,Abrams2006,Sieber2014} \\
1D periodic 
  & Both top-hat ($G(x)= 1/2r, |x|\le r$, 0 elsewhere) and exponential ($G(x) \propto \exp(-k|x|)$)
  & Attractive and repulsive coupling
  & \cite{Maistrenko2014} \\
1D periodic 
  & $G(x) \propto \exp(-|x|)$
  & Ginzburg-Landau oscillators
  & \cite{Kawamura2007,Sethia2013} \\
1D periodic
  & $G(x) \propto \exp(-|x|)$
  & Stuart-Landau oscillators 
  & \cite{Bordyugov2010} \\
1D periodic
  & Top-hat
  & Stuart-Landau oscillators 
  & \cite{Zakharova2014} \\
1D periodic 
  & Top-hat
  & 
  & \cite{Wolfrum2011_1,Wolfrum2011_2,Omelchenko2010} \\
1D periodic 
  & $G(x) \propto 1+A\cos(x)$
  & Random link removal
  & \cite{Yao2013} \\
1D periodic 
  & Arbitrary
  & Existence and stability
  & \cite{Omelchenko2013} \\
1D periodic 
  & Top-hat
  & Return map (logistic) and R\"{o}ssler system
  & \cite{Omelchenko2011} \\
1D periodic 
  & $G(x) \propto \exp(-\alpha |x|)$ or $G(x) \propto \exp(-\alpha_1 |x|) + c \exp(-\alpha_2 |x|)$
  & Hodgkin-Huxley neurons
  & \cite{Sakaguchi2006} \\
1D periodic 
  & Top-hat
  & FitzHugh-Nagumo neurons
  & \cite{Omelchenko2013_2} \\
1D line segment
  & $G(x) = \frac{1+A\cos(x)}{2k+2A\sin(k)}$
  & 
  & \cite{Zhu2012} \\
1D line segment
  & Mean-field
  & Stuart-Landau oscillators with imposed stimulation profile, delay
  & \cite{Omelchenko2008} \\
1D line
  & $G(x) \propto A e^{-a|x|}-e^{-|x|}$
  & Lighthouse model neurons
  & \cite{Chow2006} \\
1D and 2D finite size
  & $G(x) \propto \exp(-|x|)$ and $G(r) \propto K_0(r)$
  & Nonzero time delay
  & \cite{Laing2011} \\
1D periodic, 2D plane
  & $G(x) \propto 1+A\cos(x)$, $G(r) \propto K_0(r)$, power-law distribution
  & Also examined delay, heterogeneous frequencies
  & \cite{Laing2009_2} \\
1D and 2D periodic
  & Top-hat
  & Experiment, return map
  & \cite{Hagerstrom2012} \\
2D plane
  & $G(r) \propto K_0(r/r_0)$
  & Kuramoto, Ginzburg-Landau, FitzHugh-Nagumo oscillators
  & \cite{Kuramoto2003_2,Shima2004} \\
2D plane
  & $G(r) \propto \exp(-r^2)$
  & Solvable
  & \cite{Martens2009} \\
2D plane
  & Top-hat (radius $R$)
  & 
  & \cite{Kim2004,Omelchenko2012} \\	  
2D plane
  & $G(r) \propto \exp(-r)$
  & R\"{o}ssler system
  & \cite{Gu2013} \\
2D periodic (torus)
  & $G(u,v,u',v') \propto 1+\kappa\cos(u-u')+\kappa\cos(v-v')$
  & 2D analogue to \cite{Abrams2006}
  & \cite{Panaggio2013} \\
2D periodic (sphere)
  & $G(\mathbf{r},\mathbf{r'}) \propto \exp(\kappa \mathbf{r} \cdot \mathbf{r'})$
  & Both spots and spirals
  & \cite{Panaggio2014,Panaggio2014thesis} \\
Network
  & $G_{ij} \propto \exp(-\kappa d_{ij})$
  & Erd\"{o}s-R\'{e}nyi and scale-free
  & \cite{Zhu2014} \\
\end{longtable}
\end{center}
%\end{landscape}

%\bibliographystyle{unsrt}
\bibliographystyle{unsrtnat}
\bibliography{Chimera_States_Review_v4}{}
%\AtEveryBibitem{\clearfield{month}}

\end{document}